# Automatic generation of complementary auxiliary basis sets (CABS) for explicitly correlated methods.

Emmanouil Semidalas and Jan M.L. Martin

Department of Molecular Chemistry and Materials Science, Weizmann Institute of Science, 7610001 Reḥovot, Israel. Email: gershom@weizmann.ac.il

**Abstract**

Explicitly correlated calculations, aside from the orbital basis set, typically require three auxiliary basis sets: JK (Coulomb-exchange fitting), RI-MP2 (resolution of the identity MP2), and CABS (complementary auxiliary basis set). If unavailable for the orbital basis set and chemical elements of interest, the first two can be auto-generated on the fly using existing algorithms, but not the third. In this paper, we present a quite simple algorithm named autoCABS; a Python implementation under a free software license is offered at Github. For the cc-pVnZ-F12 (n=D,T,Q,5), the W4-08 thermochemical benchmark, and the HFREQ2014 set of harmonic frequencies, we demonstrate that autoCABS-generated CABS basis sets are comparable in quality to purpose-optimized OptRI basis sets from the literature, and that the quality difference becomes entirely negligible as *n* increases.
**Keywords:** explicitly correlated methods, complementary auxiliary basis set, automatic fitting basis set generation

## Introduction

Explicitly correlated ("R12") quantum chemistry methods[1–3] are becoming a mainstay of accurate wavefunction theory (WFT), particularly for thermochemistry (e.g.,[4] ) and for noncovalent interactions (e.g.,[5,6]). The addition to the orbital basis set (OBS) of 'geminal' terms that explicitly depend on interelectronic distances dramatically accelerates basis set convergence of correlation energies: Kutzelnigg and Morgan[7] showed that such a calculation's correlation energy asymptotically converges as $L^{-7}$ (with L the largest angular momentum in the basis set), compared to just $L^{-3}$ for orbital basis sets alone.

Out of different geminal forms, the F12 form $(1-\exp \gamma r_{12})/\gamma$ proposed by Ten-No[8,9] has become the de facto standard: for reasons of computational convenience, it is in practice expanded as a linear combination of Gaussian geminals (a technique advocated by Persson and Taylor[10]). (Different forms are discussed in Ref.[11])

As the short-range interelectronic cusp is adequately covered already by the F12 geminal, the requirements for the orbital basis set become quite different from those in a pure orbital calculation. Peterson and coworkers[12] first developed so-called cc-pVnZ-F12 (n=D,T,Q) basis sets for elements H, B–Ne, and Al–Ar, which were later supplemented by similar basis sets for Li, Be, Na, Mg,[13] cc-pV5Z-F12 basis sets for H, B–Ne,[14] and Al–Ar;[4] augmented versions of these basis sets[15], and cc-pVnZ-PP-F12 basis sets for the heavy p-block.[16] Very recently, the

cc-pVnZ-PP-F12 and the augmented and core-valence versions were reported for the group 11 and 12 elements.[17]

In addition to the orbital basis set and the specification of the geminal exponent, practical implementations of F12 methods, in codes such as MOLPRO,[18] ORCA,[19] and Turbomole,[20] entail no fewer than three auxiliary basis sets (ABS): the RI-JK ABS for the Coulomb and exchange integrals (which can be identical to the one used in a DF-SCF calculation); the RI-MP2 ABS (which can be identical to the one used in an orbital-only RI-MP2 calculation); and finally the CABS or complementary ABS,[21,22] which is specific to R12/F12 calculations.

What does one do for an orbital basis set where one or more ABSes are missing? While automated procedures exist (e.g., Refs.[23–25]) to generate RI-JK and RI-MP2 ABSes for a given orbital basis set, there is no corresponding procedure for CABS. Yousaf and Peterson[26] (YP) and Hill and Peterson[16] generated dedicated so-called "OptRI" basis sets for cc-pVnZ-F12 and cc-pVnZ-F12-PP, respectively, which are available in the basis set libraries of several QM codes.

YP's initial effort involved minimizing the MP2-F12 energy difference between the OptRI basis set and a very large, brute-force, reference CABS $RI_{ref}$. However, this procedure turned out to be numerically unstable, and so instead, they minimized a strictly positive objective function based on the B and V matrix elements that occur[27] in F12 theory:

$$\delta \text{RI} = \sum_{ij} \frac{(V_{ij,ij}^{\text{OptRI}} - V_{ij,ij}^{\text{RI,ref}})^2}{V_{ij,ij}^{\text{RI,ref}}} + \frac{(B_{ij,ij}^{\text{OptRI}} - B_{ij,ij}^{\text{RI,ref}})^2}{B_{ij,ij}^{\text{RI,ref}}}$$

where the indices i,j run over all occupied orbitals.

YP also obtained[28] OptRI basis sets for ordinary aug-cc-pVnZ basis sets, but it has been shown[4] that the latter have highly undesirable nonmonotonic convergence behavior (due to large BSSE) when applied in an F12 setting, and that true cc-pVnZ-F12 basis sets are free of these artifacts.

For basis set development, one can always resort to very large "brute force" fitting basis sets such as proposed by Hill, Peterson, Knizia, and Werner (HPKW)[29] (denoted reference-JKFit, reference-MP2Fit, and reference-OptRI here).

HPKW considered what one might call partial-wave-limit orbital basis sets (denoted here REF-d, REF-f, REF-g, and REF-h for the highest angular momentum retained) and their convergence to the basis set limit. In doing so, they also proposed an alternative smaller OptRI basis set to be used in conjunction with REF-h:

> "The CABS basis used in the construction of the RI consisted of even-tempered sets of diffuse and tight functions based on values of [even-tempered[30] sequence parameters] alpha and B extracted from the reference orbital basis sets, as well as exponents that intercalated pairs of existing uncontracted functions in the reference orbital basis where the gap was sufficiently large."

We found ourselves wondering if a similar procedure could not serve as an acceptable automated alternative to both brute-force reference-OptRI and purposely optimized OptRI basis sets.

In the paper below, we will show that this is indeed the case, and that such "autoCABSs" offer an alternative for basis sets where no optimized OptRI is available.

**Automatic CABS Generation Procedure**

In the present work, we present a proof of principle for the elements H–Ar and the cc-pVnZ-F12 basis sets;[12] as our validation set, we use the W4-08 set[31] of total atomization energies (TAEs) at the MP2-F12 level. In this work, they are generated from cc-pVnZ-F12 or aug-cc-pVnZ-F12 orbital basis sets but could also be developed from any arbitrary basis. The presented procedure is straightforward, and the algorithm is described below:

1) As a starting point, an orbital basis set has to be supplied as input, preferably in a machine-readable ORCA format. Then, its orbital exponents are grouped by angular momentum and whether they are uncontracted. All uncontracted exponents are retained, while a single contracted exponent of the lowest magnitude is considered to generate p exponents in cases like hydrogen due to a limited number of uncontracted exponents. That leads to a total of $l_n$ exponents for each $n$ subshell.

2) The selected $l_n$ exponents for each angular momentum type are then sorted in ascending order. Upon taking their geometric mean in consecutive pairs, $l_n - 1$ exponents are generated for each subshell and we shall denote this basis as autoCABS0.

3) In the special-case of an outermost subshell that has one exponent in the OBS, we take the geometric mean of the n–1 subshell's exponents after multiplying them by 1.5, so that both the autoCABS and OBS have functions of the same highest angular momentum.

4) A single tight (i.e., high-exponent) function for each angular momentum type is added (denoted by autoCABS0$^+$) by multiplying the highest exponent of that angular momentum by the ratio between it and the next highest exponent.

5) One diffuse (i.e., low-exponent) function is added (denoted by autoCABS0$^\pm$) to each subshell in the same 'even-tempered' manner, i.e., by dividing the lowest exponent already present by its ratio with the second lowest exponent.

6) One layer of CABS exponents is then generated by taking the geometric mean of each pair of generated exponents in the outer subshell.

7) An additional layer of exponents with the next higher angular momentum is added by taking the geometric mean of the generated exponents in pairs from (6). An autoCABS with an additional layer of zeta exponents will have an 'autoCABS1$^\pm$' designation, while an addition of a second layer will be called 'autoCABS2$^\pm$' in text.

8) When the OBS cardinal number is D, two additional tight p functions are added for the p-block elements in an even-tempered way by multiplying by 4 and 16 the largest tight function.

The preceding deterministic algorithm generates a hierarchy of autoCABSs in a reproducible manner. It is important to note that no optimization of any exponents for any objective function is involved, unlike YP's VnZ-F12-OptRI.

The algorithm has been implemented in a Python 3 script that is available at the URL https://github.com/msemidalas/autoCABS.git It expects a basis set in ORCA or MOLPRO format, but can easily be modified to use TURBOMOLE format instead. It generates an output in multiple formats for various F12-capable computational chemistry packages.

To illustrate the autoCABS generation process, let us provide an example of the generation of cc-pVTZ-F12/autoCABS variants for the atom of carbon. The cc-pVTZ-F12 orbital basis, (13s,7p,3d,2f)/[6s,6p,3d,2f], is used as input and we obtain an uncontracted [4s,5p,2d,1f] basis as autoCABS0. Next, adding to it a single tight function for each angular momentum type leads to a [5s,6p,3d,2f] autoCABS0$^+$ and an additional diffuse function generates autoCABS0$^\pm$ with [6s,7p,4d,3f]. Going one step further and adding one layer of exponents leads to autoCABS1$^\pm$ with [6s,7p,4d,3f,2g] and finally a second additional layer of exponents generates the autoCABS2$^\pm$ with [6s,7p,4d,3f,2g,1h].

**Computational Details**

All calculations were performed on the Chemfarm cluster of the Chemistry Faculty at the Weizmann Institute of Science using the MOLPRO 2022.1 electronic structure program.[18] Aside from the explicitly correlated MP2-F12 approach with the commonly used 3C(FIX) *Ansatz*[27,32], we also consider the CCSD-F12b method.[33,34]

As orbital sets, we consider the correlation consistent basis set families of cc-pVnZ-F12 (n=D,T,Q,5)[12–14,16] and aug-cc-pVnZ-F12 (n=D,T,Q),[15] abbreviated in this manuscript to VnZ-F12 and AVnZ-F12, respectively.

In addition, to be able to compare CABS truncation errors with *orbital* basis set incompleteness, we carried out basis set convergence calculations using the very large reference spdfgh basis set from Ref.[35]; partial wave convergence was gauged by truncation at f, g, and h functions: like in Ref. [14], we denote these truncated basis sets REF-*f*, REF-*g*, and REF-*h* sets, respectively.

Similarly, the HPKW reference-JKFit or the standard AVnZ/JKFit[36] are chosen as JKFit basis sets during the Hartree-Fock density fitting stage. As far as MP2Fit auxiliary basis sets for the RI-MP2 approximation, we use the reference-MP2Fit or the default AVnZ/MP2Fit basis.[37,38] Also, we settle upon the recommended values of the *β* geminal exponents in MP2-F12 and CCSD-F12b, which are *β* = 0.9, 1.0, 1.0, and 1.2 $a_0^{-1}$, respectively, for cc-pVDZ-F12,[29] cc-pVTZ-F12,[29] cc-pVQZ-F12,[29] and cc-pV5Z-F12,[39]. For the truncated spdfgh basis sets, REF-f, REF-g, and REF-h, we utilized a *β* value of 1.4 $a_0^{-1}$; the same value was also selected in ref. [4].

The choice of a CABS plays a central role in what follows. We employ the default CABSs available in MOLPRO, i.e., YP's cc-pVnZ-F12/OptRI (n=D,T,Q). For a V5Z-F12 basis, no such OptRI is available, but MOLPRO assigns the aug-cc-pwCV5Z/MP2Fit basis set[40] by default, which we will therefore consider. Also, we assessed Shaw and Hill's VnZ-F12/OptRI+, in which s and p functions, which were optimized to maximize in absolute value the CABS correction to the HF energy, are added on top of VnZ-F12/OptRI. The correlation consistent basis sets, i.e., VnZ-F12 and AVnZ-F12, are obtained from the BSE database[41] and the ESI of ref. [15], respectively, while the corresponding JKFit and MP2Fit sets are already available in MOLPRO.

The W4-08 (Weizmann 4-2008) dataset[31] was chosen for benchmarking purposes; it contains 99 atomization energies of species composed of first- and second-row atoms, while it also includes a wide range of systems dominated by dynamical to nondynamic correlation.

We employed the frozen core approximation in all calculations with the frozen core orbitals for the atoms: B-F: 1s, and Al-Cl: 1s2s2p.

For the REF-*L* basis sets, we applied a two-point extrapolation of the well-known (e.g. Ref. [42]) form $E_\infty = E(L) + \frac{E(L)-E(L-1)}{\left(\frac{L}{L-1}\right)^a - 1}$, where we set *a*=7 in accordance with the work of Kutzelnigg and Morgan, who showed[43] that the partial-wave basis set increments — that is basis set increments for basis sets fully saturated in the given angular momenta (such as is approximately the case for REF-$h$), of an explicitly correlated R12 (or, by extension, F12) calculation — converge as $l^{-8}$ with the angular momentum $l$, and hence the correlation energy for $l$ truncated at $L=L_{max}$ will have a basis set incompleteness error that depends on $L$ with the leading term $L^{-7}$. We denote such extrapolations REF-{f,g} or REF-{g,h}: as shown below in Table 1, REF-{f,g} and REF-{g,h} differ by just 0.02 kcal/mol RMSD, indicating that the actual basis set convergence is quite close to $L^{-7}$.

Ideally, as explained by YP, one would like to have as few near-linear dependences in the CABS basis set as possible. MOLPRO by default discards components with an overlap matrix eigenvalue of $10^{-8}$ or less: Inspection of the calculation outputs with our cc-pVQZ-F12 derived autoCABS shows that fewer than 0.15% of generated autoCABS functions were discarded, and typically fewer than 0.45% for the aug-cc-pVQZ-F12 derived autoCABS; for the smaller orbital basis sets, the entire autoCABS is retained. (All corresponding average percentages for W4-08 are listed in Table S1 of the Supporting Information)

## Results and Discussion

### A. Automatically generated CABSs from VnZ-F12 orbital basis sets

The generation procedure described above is flexible enough to provide various autoCABSs from VnZ-F12 OBSs and their composition is reported in Table S2 in the Supporting Information. Error statistics for the MP2-F12 atomization energies of the W4-08 dataset are gathered in Table 1.

Before we proceed, let us find out, for the standard orbital and OptRI basis sets, what error is introduced by using the standard JK (Coulomb-Exchange) and RI-MP2 basis sets. To this end, we compare between two sets of W4-08 total atomization energies: using the default fitting sets, and with reference-JKFit and reference-MP2Fit substituted for the default JK and RI-MP2 options. We find the RMSDs between them to be 0.143 kcal/mol for VDZ-F12, 0.035 kcal/mol for VTZ-F12, and a measly 0.018 kcal/mol for VQZ-F12. Also, a similar comparison with OptRI+ as CABS yields RMSD = 0.145 kcal/mol for n=D, but the error statistics do not change for larger cardinal numbers. These values are relevant for our research question at hand because they set a practical lower limit for the required accuracy in CABS: any further improvements would 'drown' in errors from JK and RI-MP2.

For perspective, using reference-JKFit and reference-MP2Fit on both sides of the comparison, the RMSD between VnZ-F12/OptRI and reference-OptRI is 0.230 kcal/mol for VDZ-F12, 0.055 kcal/mol for VTZ-F12, and 0.013 kcal/mol for VQZ-F12. These numbers reflect pure OptRI errors; for both VDZ-F12 and VTZ-F12, they are larger than the JK/RIMP2 errors above. Similarly, if we compare VnZ-F12/OptRI+ and reference-OptRI, we find lower RMSDs of 0.097, 0.040, and 0.014 kcal/mol for n=D,T,Q, respectively.

For the sake of completeness: using all-default fitting sets (VnZ-F12/OptRI, VnZ-F12/MP2Fit, def2/JK) on one side of the comparison and all "reference" fitting sets on the other, the RMSDs for n=D,T,Q are 0.243, 0.051, 0.019 kcal/mol, respectively. Again, VnZ-F12/OptRI+ as CABS with default fitting sets yields 0.126 and 0.053 kcal/mol for n=D and T, while the error statistics remain unchanged for the quadruple-zeta basis.

In order to eliminate 'confounding variables', for our autoCABS evaluations we will keep reference-JKFit and reference-MP2Fit fixed throughout, and evaluate RMSDs against reference-OptRI.

Let us first consider the autoCABS variants with no higher angular momenta than the orbital basis set, which we shall denote autoCABS0. Clearly, the 3 kcal/mol error for VDZ-F12 is completely unacceptable — an order of magnitude larger than for VDZ-F12/OptRI. For VTZ-F12, 0.17 kcal/mol is still 3 times larger than for OptRI, but the gap is clearly closing, and for VQZ-F12, 0.03 kcal/mol is only 1.5 times larger than the RI-MP2 and JK error from the default fitting basis sets.

Adding a layer of tight functions (denoted by autoCABS0$^+$) helps neither for VDZ-F12 nor for VTZ-F12, but cuts the error for VQZ-F12 down to 0.015 kcal/mol — comparable to the JK and RI-MP2 error sources. For V5Z-F12, it is now down to just 0.004 kcal/mol.

Adding a layer of diffuse functions (denoted by autoCABS0$^-$) instead offers no visible benefit, while statistics with both tight and diffuse function layers (denoted by autoCABS0$^\pm$) are comparable to those with only the tight layer added.

We note that the standard VTZ-F12/OptRI and VQZ-F12/OptRI contain one additional angular momentum beyond what is present in the corresponding orbital basis sets; in VDZ-F12/OptRI there are even two additional angular momenta. What happens if, on top of the tight and diffuse 'horizontal' layers in autoCABS, we add one 'vertical' layer, that is, additional angular momentum too (autoCABS1$^\pm$)? For VDZ-F12 this cuts RMSD by a factor of about six, and for VTZ-F12 by a factor of about three. For VQZ-F12 there is only a slight improvement, and for V5Z-F12 no effect is observed to three decimal places.

The second layer about halves RMSD for VDZ-F12 to 0.29 kcal/mol, not much greater than the 0.23 kcal/mol for VDZ-F12/OptRI. For VTZ-F12 the additional reduction is already becoming insignificant, from 0.070 to 0.061 kcal/mol, and for VQZ-F12 and V5Z-F12 none is seen to three decimal places.

Another possibility is to consider tight functions and a single additional layer of exponents (autoCABS1$^+$). In the case of VDZ-F12/autoCABS1$^+$, we notice a significant deterioration in RMSD by 0.81 kcal/mol relative to the corresponding autoCABS1$^\pm$ with additional diffuse functions. Error statistics worsen by 0.025 kcal/mol for VTZ-F12/autoCABS1$^+$, but there is an insignificant improvement for n=Q. The inclusion of diffuse functions was evaluated with respect to reference-OptRI as CABS, but the statistics do not materially change if one compares the two autoCABSs against either YP's OptRI or SH's OptRI+ as CABS.

There are no official OptRI basis sets for AVnZ-F12; the default choice in MOLPRO is to use VnZ-F12/OptRI for AVnZ-F12. This choice becomes plausible in light of the above observations that adding a layer of diffuse functions to the OptRI basis set has a negligible effect for VnZ-F12, but let us consider statistics relative to Reference-OptRI. The RMSD values are 0.284, 0.082, and 0.019 kcal/mol, respectively, for AVnZ-F12 (n=D,T,Q). For AVDZ-F12, one autoCABS with two extra angular momenta (autoCABS2$^\pm$) actually has a smaller RMSD = 0.249 kcal/mol, while autoCABS with one extra angular momentum (autoCABS1$^\pm$) is definitely competitive with VnZ-F12/OptRI for AVTZ-F12 and AVQZ-F12.

Table 1. RMSDs (kcal/mol) of various VnZ-F12/autoCABSs for the W4-08 dataset with MP2-F12.

| Orbital basis | CABS | RMSD[a] | RMSD[b] | RMSD[c] | RMSD[a] | RMSD[b] | RMSD[c] | RMSD[a] | RMSD[b] | RMSD[c] | RMSD[a] | RMSD[c] |
|---|---|---|---|---|---|---|---|---|---|---|---|---|
|  | RMSD relative to | OptRI | OptRI+ | REF | OptRI | OptRI+ | REF | OptRI | OptRI+ | REF | OptRI | REF |
|  | n = | D | D | D | T | T | T | Q | Q | Q | 5 | 5 |
| VnZ-F12 | autoCABS0 | 3.059 | 2.999 | 3.042 | 0.172 | 0.187 | 0.199 | 0.032 | 0.033 | 0.035 | N/A | 0.016 |
|  | autoCABS0$^+$ | 3.073 | 3.017 | 3.060 | 0.167 | 0.175 | 0.184 | 0.013 | 0.017 | 0.015 | N/A | 0.004 |
|  | autoCABS0$^-$ | 3.057 | 2.998 | 3.039 | 0.185 | 0.200 | 0.210 | 0.031 | 0.032 | 0.032 | N/A | 0.003 |
|  | autoCABS0$^\pm$ | 3.075 | 3.019 | 3.061 | 0.180 | 0.189 | 0.197 | 0.016 | 0.020 | 0.015 | N/A | 0.003 |
|  | autoCABS1$^+$ | 1.349 | 1.324 | 1.386 | 0.064 | 0.081 | 0.096 | 0.009 | 0.012 | 0.013 | N/A | 0.004 |
|  | autoCABS1$^\pm$ | 0.538 | 0.541 | 0.590 | 0.039 | 0.071 | 0.070 | 0.012 | 0.014 | 0.009 | N/A | 0.003 |
|  | autoCABS2$^\pm$ | 0.208 | 0.256 (0.238[f]) | 0.291 (0.276[f]) | 0.028 | 0.065 | 0.061 | 0.012 | 0.014 | 0.009 | N/A | N/A |
|  | YP's VnZ-F12/OptRI | REF | 0.143[d] | 0.230 | REF | 0.035[d] | 0.055 | REF | 0.018[d] | 0.013 | N/A | N/A |
|  | VnZ-F12/OptRI+ | 0.145[e] | REF | 0.097 | 0.035[e] | REF | 0.040 | 0.018[e] | REF | 0.014 | N/A | N/A |
|  | VnZ-F12/MP2Fit |  |  | 0.292 |  |  | 0.050 |  |  | 0.006 |  |  |
|  | VnZ-F12/JKFit |  |  | 0.225 |  |  | 0.027 |  |  | 0.007 |  |  |
|  | reference-OptRI | 0.243[d] | 0.126[e] | N/A | 0.051[d] | 0.053[e] | N/A | 0.019[d] | 0.019[e] | N/A | N/A | N/A |
|  | awCV5Z/MP2Fit | N/A | N/A | N/A | N/A | N/A | N/A | N/A | N/A | N/A | 0.004[d] | 0.001 |
|  | RMSD relative to | L$^{-7}$ REF-{g,h} |  |  |  |  |  |  |  |  |  |  |
| VnZ-F12 | VnZ-F12/MP2Fit | 0.886 |  |  | 0.294 |  |  | 0.083 |  |  |  |  |
|  | VnZ-F12/JKFit | 0.858 |  |  | 0.282 |  |  | 0.081 |  |  |  |  |
|  | VnZ-F12/OptRI | 1.018 |  |  | 0.293 |  |  | 0.083 |  |  |  |  |
|  | VnZ-F12/OptRI+ | 0.921 |  |  | 0.276 |  |  | 0.085 |  |  |  |  |
|  | reference-OptRI | 0.947 |  | REF | 0.284 |  | REF | 0.080 |  | REF | 0.033 | REF |
| REF-f | ditto | 0.329 |  |  |  |  |  |  |  |  |  |  |
| REF-g | ditto | 0.062 |  |  |  |  |  |  |  |  |  |  |
| REF-h | ditto | 0.013 |  |  |  |  |  |  |  |  |  |  |
| L$^{-7}$ REF-{f,g} | ditto | 0.025 |  |  |  |  |  |  |  |  |  |  |
| L$^{-7}$ REF-{g,h} | ditto | REF |  |  |  |  |  |  |  |  |  |  |

autoCABS variants that are generated from VnZ-F12 basis sets against: (a) YP's VnZ-F12-OptRI (n=D,T,Q) as CABSs; (b) VnZ-F12-OptRI+ (n=D,T,Q) as CABSs; (c) reference-OptRI as a CABS. As JKFit and MP2Fit basis sets, the reference ones are chosen (a-c). RMSDs with respect to: (d) YP's VnZ-F12-OptRI (n=D,T,Q) as CABSs with VnZ-F12/JK and VnZ-F12/MP2Fit as JKFit and MP2Fit, respectively.
For n=5, AWCV5Z/MP2Fit was chosen as CABS, but V5Z-F12/JK and V5Z-F12/MP2FIT as JK and MP2Fit sets; (e) VnZ-F12-OptRI+ (n=D,T,Q) as CABS with the default JKFit and MP2Fit as in (d); (f) VDZ-F12/autoCABS2$^\pm$ with two tight p functions for the p-block elements.

Another angle from which we can consider the significance of fitting basis set errors is to compare them with basis set incompleteness errors. Our reference here is REF-{g,h} with all three fitting basis sets "reference" quality; the RMSD of REF-h is equivalent to the RMS extrapolation size, and is just 0.013 kcal/mol; REF-{f,g} has RMSD = 0.025 kcal/mol, indicating that we are indeed near the basis set limit. (Incidentally, REF-h with the reduced OptRI basis set given in the ESI of Hill et al. (2009) introduces an RMSD = 0.007 kcal/mol.)

What happens for VnZ-F12? In order to eliminate confounding factors, we again use all three "reference" fitting basis sets. VnZ-F12 RMSD = 0.947, 0.284, 0.080, and 0.033 kcal/mol for n=D, T,Q,5, respectively, all several times larger than the autoCABS fitting basis set error. For AVnZ-F12 we have RMSD = 0.901, 0.259, and 0.075 kcal/mol for n=D,T,Q, respectively: again, several times over the autoCABS errors.

Next, supplementing the quadruple-zeta autoCABS with an additional layer of exponents, an insignificant lowering of the RMSD from 0.016 to 0.012 kcal/mol occurs, while a second layer (autoCABS2$^\pm$) yields similar performance.

Regarding the choice of a CABS with the V5Z-F12 orbital basis, the RMSD statistics are available for our autoCABSs against the reference-OptRI as CABS. For V5Z-F12/autoCABS, one of the lowest RMSD values of 0.003 kcal/mol is obtained by including diffuse and tight exponents, while the error statistics do not materially change after adding an outer layer of exponents. The MOLPRO default for V5Z-F12 is to use awCV5Z/MP2FIT as the CABS, and this expedient yields the lowest error (0.001 kcal/mol) against reference-OptRI.

Furthermore, with the truncated REF-f, REF-g, and REF-h as orbital basis sets, while considering as CABS the reference-OptRI, we obtain RMSDs of 0.329, 0.062, and 0.013 kcal/mol, respectively, relative to the extrapolated $L^{-7}$ REF-{g,h}.

What about the energetics based on SH's VnZ-F12/OptRI+ as CABS? Our larger autoCABSs yield slightly higher RMSDs than in the previous comparison against YP's VnZ-F12/OptRI, but the errors remain lower than those vs. reference-OptRI. For example, the largest double-zeta autoCABS with tight and diffuse exponents plus two additional layers of exponents (autoCABS2$^\pm$) reaches RMSD = 0.256 kcal/mol and 0.291 kcal/mol against OptRI+ and reference-OptRI, respectively. Overall, the error statistics are still unacceptably large for n=D. However, the RMSDs are below 0.1 kcal/mol for n=T, even with an autoCABS that contains tight exponents and a single additional layer of exponents, and there is a further improvement for n=Q. The RMSDs steeply decline toward the basis set limit, and similar conclusions can be drawn to the YP's OptRI case.

If, inspired by the construction of the OptRI+ sets[44], we add to our largest autoCABS2$^\pm$ one pair of additional tight p functions for the p-block elements, we find for n=D the RMSD to drop from 0.256 to 0.238 kcal/mol with respect to VDZ-F12/OptRI+ as CABS. Also, there is a similar improvement in statistics against the reference-OptRI as CABS by just 0.015 kcal/mol.

For the sake of comparison, we also check the recommendation of using VnZ-F12/MP2Fit or VnZ-F12/JKFit sets as CABS in MP2-F12. In that scenario, there is some deterioration for the double zeta basis, where the VDZ-F12/MP2Fit set underperforms OptRI by 0.062 kcal/mol, but for n=T and Q, the statistics agree within a few hundredths of 1 kcal/mol. On the other hand, VDZ-F12/autoCABS requires tight and diffuse functions and two additional layers of exponents to perform better by 0.036 kcal/mol over VDZ-F12/MP2Fit.

Table 2. Number of basis functions across various CABSs.

| Species | H$_2$ | | | N$_2$ | | | P$_2$ | | |
|---|---|---|---|---|---|---|---|---|---|
| CABS        n= | D[a] | T | Q | D[a] | T | Q | D[a] | T | Q |
| VnZ-F12/autoCABS2$^{\pm}$ | 60 | 152 | 172 | 134 | 194 | 266 | 176 | 204 | 276 |
| OptRI | 44 | 82 | 100 | 132 | 150 | 172 | 132 | 150 | 172 |
| OptRI+ | 48 | 86 | 104 | 146 | 164 | 186 | 146 | 164 | 186 |
| VnZ-F12/MP2Fit | 46 | 92 | 160 | 144 | 212 | 336 | 184 | 294 | 388 |
| VnZ-F12/JKFit | 64 | 92 | 152 | 172 | 208 | 284 | 256 | 292 | 368 |
| reference-OptRI | | 822 | | | 1004 | | | 1124 | |

(a) VDZ-F12/autoCABS2$^{\pm}$ with two tight p functions for the p-block elements

Moreover, with a JKFit as CABS, we find that VDZ-F12/JKFit outperforms the corresponding MP2Fit by 30%, and for the triple zeta basis, there is 0.023 kcal/mol amelioration, while for n=Q the RMSDs match. Even though neither the MP2Fit nor the JKFit basis sets were originally constructed for explicitly correlated methods, their good performance could be attributed to their large size, thus further increasing the computational effort. As a way of illustration, the number of RI functions for a few diatomics can be found for several CABSs in Table 2.

Overall, the automatically generated autoCABSs display the same 'correlation consistent' systematic decay of RMSD with increasing n as YP's VnZ-F12-OptRI ones. However, to obtain a similar performance, an additional layer of exponents is needed, at least for n=D. A somewhat inferior performance is actually observed for n=D, where an autoCABS2$^{\pm}$ is 0.12 kcal/mol inferior to VDZ-F12-OptRI. However, as n increases, any autoCABS1 approaches the performance of YP's VnZ-F12-OptRI. In fact, for a quintuple autoCABS, the W4-08 energetics are reproduced within a few hundredths of 1 kcal/mol relative to the brute-force reference-OptRI. These results highlight the accurate performance of the hierarchical structure obtained from the generation procedure.

## B. Automatically generated CABSs from AVnZ-F12 orbital basis sets

To benchmark the autoCABSs generated from the AVnZ-F12 orbital basis sets against the reference-OptRI as CABS, we calculate the associated error statistics for W4-08 in Table 3. The composition of those autoCABSs can be found in Table S3 in the Supporting Information.

The lowest errors are recorded for AVQZ-F12/autoCABS0$^{\pm}$ and its variants to the reference-OptRI as CABS. Adding either one or two layers of exponents to the AVQZ-F12/autoCABS0$^{\pm}$ brings down the RMSD from 0.020 to 0.016 kcal/mol. For perspective, the corresponding VQZ-F12/autoCABS0$^{\pm}$ with a VQZ-F12 as OBS, performs identically with RMSD = 0.015 kcal/mol (see Table 1). Also, AVTZ-F12/autoCABS0$^{\pm}$ reaches RMSD = 0.200 kcal/mol, and one additional layer of exponents reduces error statistics by almost half, while a second layer yields only a 0.002 kcal/mol improvement. However, the AVDZ-F12/autoCABS0$^{\pm}$ shows a performance that we deem unacceptably poor at 2.960 kcal/mol. Thus, the additional angular momentum layers are needed; with one or two such layers, the recorded RMSDs plunge to 0.386 and 0.249 kcal/mol, respectively.

The performance of VDZ-F12/OptRI+ is better by 0.155 kcal/mol over YP's VDZ-F12/OptRI as CABS and when an AVDZ-F12 basis is used as OBS. For larger cardinal numbers, the gap between OptRI+ and OptRI is further reduced by 0.023 and (marginally) 0.002 kcal/mol for n=T and Q, respectively.

What happens if we substitute the smaller VnZ-F12/autoCABSs for the AVnZ-F12/autoCABSs in conjunction with the largest AVnZ-F12 as orbital basis sets? When using smaller CABSs, such as the VDZ-F12/autoCABS1$\pm$ variant, the RMSDs increase by 50% vs. AVDZ-F12/autoCABS1$\pm$, while for two additional layers of exponents, this error becomes 1.4 times larger. However, for an AVTZ-F12 orbital basis, either VTZ-F12- or AVTZ-F12-derived CABSs yield identical statistics, and the same applies to n=Q.

Table 3. RMSDs (kcal/mol) of various autoCABSs evaluated against the large reference-OptRI for the W4-08 dataset with MP2-F12.

| Orbital basis | CABS | RMSD[a] | RMSD[b] | RMSD[a] | RMSD[b] | RMSD[a] | RMSD[b] |
|---|---|---|---|---|---|---|---|
| | n= | D | D | T | T | Q | Q |
| AVnZ-F12 | autoCABS0$\pm$ | 2.960 | 2.994 | 0.200 | 0.204 | 0.020 | 0.021 |
| | autoCABS1$\pm$ | 0.386 | 0.578 | 0.098 | 0.099 | 0.016 | 0.016 |
| | autoCABS2$\pm$ | 0.249 | 0.311 | 0.096 | 0.093 | 0.016 | 0.016 |
| | YP's VnZ-F12/OptRI | 0.284 | | 0.082 | | 0.019 | |
| | VnZ-F12/OptRI+ | 0.129 | | 0.059 | | 0.017 | |
| REF-f | reference-OptRI | 0.328[c] | | | | | |
| REF-g | ditto | 0.062[c] | | | | | |
| REF-h | ditto | 0.013[c] | | | | | |
| $L^{-7}$ REF-{f,g} | ditto | 0.025[c] | | | | | |

(a) autoCABS variants are generated from AVnZ-F12 basis sets (n=D,T,Q).
(b) As (a), but all autoCABS variants are generated from non-augmented VnZ-F12 basis sets.
(c) RMSDs are calculated against the extrapolated reference $L^{-7}$ REF-{g,h}.

With the $L^{-7}$ extrapolated REF-{g,h} as a reference and the truncated spdfgh basis sets as orbital basis sets, we find that the RMSDs are gradually reduced from REF-f to REF-g (0.328 to 0.062 kcal/mol), and even further with REF-h (RMSD = 0.013 kcal/mol). The $L^{-7}$ REF-{f,g} extrapolation is 0.012 kcal/mol higher compared to REF-h as orbital basis.

**C. Performance of autoCABS in CCSD-F12b**

While MP2-F12 can be quite powerful in some contexts (particularly noncovalent interactions, e.g.,[5,6,45,46]) in conjunction with a CCSD(T) – MP2 correction in a small to medium basis set, for thermochemical applications one would like to employ an explicitly correlated coupled cluster method. As the (T) step does not benefit from F12,[47] we have focused here on the widely used CCSD-F12b method.[33,34] for the W4-08 total atomization energies, using the cc-pVnZ-F12 basis set. As the reference for each basis set, we use reference-OptRI as the CABS. In order to reduce confounding factors, reference-JKFit and reference-MP2Fit were otherwise used throughout. Error statistics are presented in Table 4.

Irrespective of the CABS family chosen — VnZ-F12/OptRI, VnZ-F12/OptRI+, VnZ-F12/MP2Fit, VnZ-F12/JKFit, or autoCABS — the differences with reference-OptRI rapidly decay with $n$. The RMSDs follow a steep decline towards the basis set limit (Table 4) against the large reference-OptRI as CABS; a similar error reduction is seen for the MP2-F12 statistics (Table 1). For the *recommended* autoCABSs, the RMSD errors are only larger than those for YP's VnZ-F12/OptRI by 0.066 and 0.014 kcal/mol for n=D and T, respectively, while the improvement in the quadruple autoCABS is just 0.004 kcal/mol. The term 'recommended' refers to the largest generated basis sets that include tight and diffuse functions, two additional angular momenta, and two additional tight p functions for the p-block elements for n=D, i.e. autoCABS2$^{\pm}$, while for n=T and Q, we consider just one additional angular momentum and no tight p functions on the p-block elements (autoCABS1$^{\pm}$). The VnZ-F12 (n=D,T,Q) are chosen as orbital basis sets as previously, while the reference sets are used for the JK and MP2 fitting steps in following comparisons.

The OptRI+ fitting sets[44], which add high-exponent functions to OptRI in order to improve the CABS correction to the HF component, very significantly reduce RMSD for n=D, but much less so for n=T and insignificantly for n=Q, where all options perform equally well. Taking that into consideration, we attempted adding tight p functions to the VDZ-F12/autoCABS for all p-block elements (similarly to the OptRI+ sets[44]), but we found a marginal improvement relative to the 'non-extended' VDZ-F12/autoCABS by 0.012 kcal/mol for one additional tight p function and 0.017 kcal/mol for a pair of them for the p-block elements.

**D. Repurposing RI-JK or RI-MP2 fitting basis set as CABS; computational cost considerations**

A workaround sometimes used when no CABS is available for a given orbital basis set is to repurpose any available MP2Fit or JKFit basis set as the CABS. How well or poorly does this work for the test set at hand? For n=D and n=T, JKFit-as-CABS is definitely superior to MP2Fit-as-CABS and autoCABS, while for n=Q all three RMSDs are under 0.01 kcal/mol (see the middle block of Table 4).

But in order to put the extra "CABS error" into perspective, it ought to be compared with the basis set incompleteness error. We evaluated this by comparing computed total atomization energies with those obtained previously[48] at the CCSD-F12b/aug-cc-pwCV5Z level, which had previously been shown[14] to be within about 0.01 kcal/mol RMSD from the basis set limit. The RMSD for CCSD-F12b/VnZ-F12 atomization energies is given at the bottom of Table 4. It is thus indeed seen that, even for VDZ-F12, the error incurred by using autoCABS is an order of magnitude smaller than the basis set incompleteness error, and that this scale difference is even more pronounced for the larger basis sets. We hence conclude that the additional CABS error is insignificant in the larger scheme of things.

Table 4. RMSDs (kcal/mol) with a reference-OptRI as CABS in CCSD-F12b.

| Orbital basis | CABS | n= | RMSDs from CCSD-F12b/VnZ-F12 | | |
|---|---|---|---|---|---|
| | | | D | T | Q |
| VnZ-F12 | autoCABS1$^{\pm}$ | | 0.299 | 0.069 | 0.008 |
| | autoCABS2$^{\pm a}$ | | 0.287 | | |
| | autoCABS2$^{\pm b}$ | | 0.282 | | |
| | YP's VnZ-F12/OptRI | | 0.233 | 0.055 | 0.012 |
| | VnZ-F12/OptRI+ | | 0.092 | 0.039 | 0.014 |
| | VnZ-F12/MP2Fit | | 0.295 | 0.050 | 0.006 |
| | VnZ-F12/JKFit | | 0.226 | 0.027 | 0.007 |
| | RMSDs from CCSD-F12b/awCV5Z[48] | | | | |
| VnZ-F12 | YP's VnZ-F12/OptRI | | 2.277 | 0.737 | 0.179 (0.065$^c$) |
| AVnZ-F12 | | | | 0.730 | 0.175 |

Preferred choice of autoCABS that include (a) one additional tight p function or (b) two additional p functions for the p-block elements. (c) V{T,Q}Z-F12 with an extrapolation coefficient of 4.596 given in Table X in ref. [29].

What about computational cost? In order to assess this, we measured wall clock times for some aromatic molecules, ranging from benzene to bithiophene. In all these calculations, identical hardware was used, each job ran by itself on an otherwise empty node with two 8-core Intel Xeon E5-2630 CPUs and all I/O operations are performed on a local 3.6 TB SSD RAID array with a bandwidth of about 3 GB/s. In order to reduce the effect of minor clock speed fluctuations due to environmental factors such as temperature inhomogeneity in the server rack the wall times are averaged over four different runs for each species (Table 5). Wall clock times for autoCABS, OptRI, OptRI+ are all comparable, as are (somewhat surprisingly) JKFit-as-CABS; all these options represent a dramatic speedup over reference-OptRI for VDZ-F12, and still a significant one for VTZ-F12, but by the time one gets to VQZ-F12, the additional cost of reference-OptRI has become quite modest in practice, e.g., 2.3h out of a total of 37.1h for bithiophene.

Table 5. Mean wall clock times (min) for various CABSs in CCSD-F12b on two 8-core Intel Xeon E5-2630 CPUs (2.40 GHz)

| CABS | VnZ-F12/ autoCABS | | | VnZ-F12/ OptRI | | | VnZ-F12/ OptRI+ | | | VnZ-F12/ MP2Fit | | | VnZ-F12/ JKFit | | | reference-OptRI | | |
|---|---|---|---|---|---|---|---|---|---|---|---|---|---|---|---|---|---|---|
| Species n= | D | T | Q | D | T | Q | D | T | Q | D | T | Q | D | T | Q | D | T | Q |
| benzene | 3 | 19 | 167 | 3 | 19 | 164 | 3 | 19 | 164 | 3 | 19 | 169 | 3 | 19 | 168 | 19 | 37 | 186 |
| phenol | 5 | 41 | 334 | 5 | 41 | 339 | 5 | 41 | 343 | 5 | 41 | 342 | 5 | 40 | 341 | 33 | 71 | 382 |
| phenyl-phosphine | 7 | 50 | 449 | 7 | 50 | 447 | 7 | 50 | 457 | 7 | 50 | 454 | 7 | 50 | 456 | 42 | 89 | 493 |
| bithiophene | 31 | 239 | 2087 | 31 | 238 | 2059 | 31 | 239 | 2078 | 31 | 240 | 2083 | 31 | 239 | 2059 | 138 | 356 | 2223 |

As orbital basis sets, the corresponding VnZ-F12 are employed and all calculations are carried out in C$_1$ symmetry. All reported wall clock times are averages over four different runs. For the autoCABSs, we consider the recommended variants of autoCABS1$^{\pm}$ for n = T and Q, and VDZ-F12/autoCABS2$^{\pm}$ with a pair of tight p functions for the p-block elements.

How much does the CABS selection affect the timings for the individual steps in CCSD-F12b? For illustration purposes, we limit ourselves to the molecules listed in Table 5, and for various VnZ-F12 orbital basis sets, the average contribution (%) to the total wall clock times is examined (see Table S4 in ESI). We also consider the default VnZ-F12/JKFit and VnZ-F12/MP2Fit basis sets for the MP2Fit and JKFit parts, respectively. Overall, the average contribution to the wall clock times depends strongly on the orbital basis set choice, and the variability in allocated time is smaller across various CABSs, except for the reference-OptRI. The lion's share of wall clock time is spent on the CCSD iterations for CABSs with n=D. Towards larger cardinal numbers, however, we find the 3-index transformation in the F12 part to have a higher contribution, as much as one-third of the total wall clock time, while the RI-MP2-F12 step seemingly has a minimum contribution. The one notable exception is when the reference-OptRI is used as CABS; the RI-MP2-F12 step then represents a significant portion of total wall time for smaller orbital basis sets, with 82.9% for n=D and 42.5% for n=T, respectively, while for n=Q, the 3-index transformation in the F12 part and the CCSD iterations are the dominant steps. Also, the time spent on the F12b part is always a tiny fraction of the total.

**E. Vibrational frequencies**

It has previously been shown by Rauhut et al.[49] that the rapid basis set convergence of F12 methods also extends to vibrational frequencies. In fact, using the CCSD(T)(F12*) approach[50] (known as CCSD(T)-F12c to MOLPRO users) Kesharwani and Martin found[51] for the HFREQ2014 dataset of small polyatomics,[51] that even the modest cc-pVDZ-F12 basis set deviates by just 3 cm$^{-1}$ RMS from the basis set limit, i.e., just over half the intrinsic error of CCSD(T) itself.[51] F12 approaches have been used (e.g., Refs.[52,53]) to fill experimental data lacunae, particularly for astrochemically important molecules.

A reviewer suggested that we consider whether our various autoCABSs cause any numerical errors in the harmonic frequencies. Shaw and Hill[44] (see Table 6 in that reference) compared cc-pVnZ-F12/OptRI and cc-pVnZ-F12/OptRI+ to the Reference-OptRI for a set of 27 diatomic molecules: mean absolute deviations (MADs) were {3.6, 1.9, 0.7} cm$^{-1}$ for OptRI with n={D,T,Q}, and {3.3, 1.8, 0.7} cm$^{-1}$ for OptRI+.

We chose to consider instead the HFREQ2014 set as being somewhat more representative of practical applications. Unfortunately, many of our frequency calculations with the Reference-OptRI CABS met with failure due to numerical problems: therefore, instead, in Table 6 in the present work we offer statistics of discrepancies between autoCABS and OptRI as well as OptRI+. Concerning the accuracy thresholds, we use $10^{-12}$ $E_h$ for the energy, $10^{-18}$ for the two-electron integrals, and $10^{-20}$ for the prefactor in two-electron integrals. The geometry optimization is terminated if the gradient falls below $10^{-6}$ (in addition to the energy criterion). Also, following Ref.[51], we set thrcabs=$10^{-9}$, thrcabs_rel=$10^{-9}$, and ortho_cabs=1, and used a step size of 0.005 au in the numerical differentiation for the force constants. As JKFit and MP2Fit sets, we use the large reference-JKFit and reference-MP2Fit.

Table 6. Deviations of harmonic frequencies (cm$^{-1}$) using automatically generated autoCABSs relative to either YP's VnZ-F12/OptRI or SH's VnZ-F12/OptRI+ as CABSs.

| | CCSD(F12*) harmonic frequencies | | | | | |
|---|---|---|---|---|---|---|
| | relative to cc-pVnZ-F12/OptRI | | | relative to cc-pVnZ-F12/OptRI+ | | |
| CABS | VDZ-F12/ autoCABS2$^\pm$ | VTZ-F12/ autoCABS1$^\pm$ | VQZ-F12/ autoCABS1$^\pm$ | VDZ-F12/ autoCABS2$^\pm$ | VTZ-F12/ autoCABS1$^\pm$ | VQZ-F12/ autoCABS1$^\pm$ |
| MSDiff | -0.32 | 0.00 | -0.14[a] | -1.04 | -0.19 | -0.14[a] |
| MAD | 1.43 | 0.22 | 0.23[a] | 1.69 | 0.34 | 0.26[a] |
| RMSD | 2.84 | 0.32 | 0.46[a] | 2.98 | 0.44 | 0.49[a] |
| Max+Dev | 4.24 | 1.37 | 0.69 | 4.18 | 1.51 | 0.98 |
| Max-Dev | -18.12[c] | -1.01 | -3.32[b] | -17.42[c] | -1.12 | -3.56[b] |
| | CCSD(T)(F12*) harmonic frequencies | | | | | |
| | relative to cc-pVnZ-F12/OptRI | | | relative to cc-pVnZ-F12/OptRI+ | | |
| CABS | VDZ-F12/ autoCABS2$^\pm$ | VTZ-F12/ autoCABS1$^\pm$ | VQZ-F12/ autoCABS1$^\pm$ | VDZ-F12/ autoCABS2$^\pm$ | VTZ-F12/ autoCABS1$^\pm$ | VQZ-F12/ autoCABS1$^\pm$ |
| MSDiff | -0.31 | 0.01 | -0.07 | -1.10 | -0.23 | -0.07 |
| MAD | 1.42 | 0.23 | 0.18 | 1.73 | 0.40 | 0.19 |
| RMSD | 2.90 | 0.35 | 0.28 | 3.10 | 0.53 | 0.28 |
| Max+Dev | 4.18 | 1.70 | 0.81 | 4.06 | 1.73 | 0.76 |
| Max-Dev | -18.83[c] | -0.94 | -1.29 | -18.46[c] | -1.73 | -1.30 |

(a) $C_2H_2$ is excluded from the statistics because of numerical instabilities. (b) torsion mode of $CH_3OH$; next largest difference just 1.2 cm$^{-1}$. (c) $\pi_g$ bending mode of $C_2H_2$; next largest negative difference -7.7 and 7.6 cm$^{-1}$, respectively, for $CS_2$ bend.

It is readily seen that the MADs between our autoCABS varieties and OptRI viz. OptRI+ are actually considerably smaller than what was found by Shaw and Hill[44] between OptRI/OptRI+ and reference-OptRI for diatomics. This holds true regardless of whether we consider CCSD(F12*) or CCSD(T)(F12*).

For CCSD(F12*)/cc-pVQZ-F12, as well as for both CCSD(F12*) and CCSD(T)(F12*) with the cc-pVDZ-F12 basis sets, RMSDs (root mean square differences) are considerably larger than the $\approx$(5/4)MAD for a normal distribution,[54] which indicates the presence of outliers. For the cc-pVQZ-F12 case, it turns out to be a 3.6 cm$^{-1}$ error in the internal rotation mode of methanol, which actually goes away at the CCSD(T)(F12*) level with the same basis set. For VDZ-F12, the culprit is the $\pi_g$ bending mode of acetylene: this is a well-documented case, see Ref.[55] and references therein, of hypersensitivity to the basis set due to internal basis set superposition error. (A more extreme version of this problem is seen for the out-of-plane bending modes of benzene: see, e.g., Refs.[56–58] and references therein.)

For perspective, using as CABS VDZ-F12/OptRI, AVQZ/JKFit, and AV5Z/JKFit, we obtained, respectively, 619.0, 614.3, and 633.2 cm$^{-1}$. (As an aside, we evaluated the harmonic frequencies of acetylene for the cc-pVDZ-F12 orbital basis set with all of our automatically generated VDZ-F12/autoCABS variants at CCSD(T)-F12b, CCSD(T)(F12*), and MP2-F12 levels (see Tables S5-S7 in the Supporting Information). Perhaps owing to a felicitous error compensation, using the smaller VDZ-F12/ autoCABS1$^\pm$, discrepancies with cc-pVDZ-F12/OptRI are remarkably lowered to 1.67, 4.55, and 1.96 cm$^{-1}$ for CCSD(T)(F12*), CCSD(T)-F12b, and MP2-F12, respectively.) We would like to argue that the real culprit here is not the CABS but the inadequate orbital basis set, which should not be used for a system like acetylene in the first place.

**Conclusions**

We have presented an automatic procedure to generate complementary auxiliary basis sets (autoCABSs) from arbitrary orbital basis sets, and have tested it at the MP2-F12 and CCSD-F12b levels for the W4-08 thermochemical benchmark and the F12-correlation consistent cc-pVnZ-F12 and aug-cc-VnZ-F12 basis sets (n=D,T,Q,5). Performance was compared to Yousaf and Peterson's VnZ-F12-OptRI, Shaw and Hill's VnZ-F12/OptRI+, and to the large reference RI basis set of HPKW.[29] For perspective, MP2-F12/CBS results obtained from *spdfgh* and *spdfgh* truncations of the large orbital reference basis set from HPKW were also considered. We were able to draw the following conclusions:

- The generated autoCABSs from VnZ-F12 orbital basis sets reproduce the accuracy of YP's VnZ-F12-OptRI and SH's VnZ-F12-OptRI+ as long as: (1) at least one tight function per shell is added; (2) a single additional angular momentum layer of exponents is considered for n=T (it can be omitted for n=Q and especially n=5); and (3) two additional angular momentum layers are added for n=D (autoCABS2$_-^+$).
- For the V5Z-F12 orbital basis set, both our generated V5Z-F12/autoCABS and awCV5Z/MP2Fit used as CABS yield deviations below 0.005 kcal/mol from the large reference CABS.
- AVnZ-F12, by and large, exhibits the same trend as VnZ-F12. The MOLPRO practice to use unmodified VnZ-F12/OptRI as CABS is vindicated for n=T and higher; for n=D adding a diffuse layer to the CABS might have been advisable, but the error incurred by omitting it is outweighed by other error sources such as basis set incompleteness.
- The CCSD-F12b results indicate similar error trends with the MP2-F12 ones and the generated autoCABSs reproduce accurately the energetics towards the basis set limit.
- The practice of repurposing MP2Fit and especially JKfit basis sets as CABS appears to be justified from an accuracy point of view, although autoCABS basis sets will be more compact and hence economical, albeit still somewhat larger than purpose-optimized basis sets like OptRI and OptRI+.
- Performance of autoCABSs for harmonic frequencies was likewise found to be satisfactory.
- The proposed simple algorithm provides on-the-fly complementary-auxiliary-basis-sets with a consistent convergence behavior and it does not require the traditional optimization of individual exponents per atom.
- This algorithm may facilitate explicitly correlated calculations for heavier elements (particularly transition metals), as well as development of specialized basis sets for these.

**Supplementary material**

Total electronic energies, calculated energetics, and associated statistics in the W4-08 dataset; calculated frequencies for the HFREQ2014 species; average percentage of discarded functions for the W4-08 species in MP2-F12; timing comparisons for individual steps; composition of autoCABSs generated from VnZ-F12 and AVnZ-F12 orbital basis sets; average contribution to the total wall clock times in CCSD-F12b of four selected aromatics.

Furthermore, the autoCABS program is available for download from https://github.com/msemidalas/autoCABS.git and a brief documentation file has been included.

**Funding**

This research was supported in part by the Israel Science Foundation (grant 1969/20) and by the Minerva Foundation (grant 2020/05). The work of E.S. on this scientific paper was supported by the Onassis Foundation–Scholarship ID: FZP 052-2/2021-2022.

**TABLE OF CONTENTS GRAPHIC**

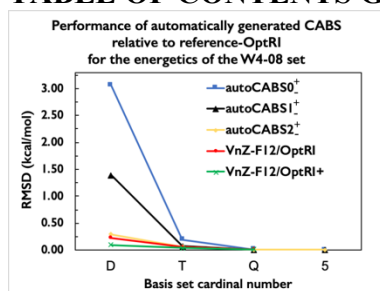

**75-WORD SUMMARY FOR TABLE OF CONTENTS**

Explicitly correlated methods exhibit dramatically faster basis set convergence compared to orbital-only methods. The lack of an appropriate CABS (complementary auxiliary basis set) can be an obstacle to application for less-common chemical elements or orbital basis sets. We propose a simple automated "autoCABS" approach to generate CABSes from an arbitrary basis set; for test sets where purpose-built CABSes are available, we show that they are comparable in accuracy and compactness to those generated by autoCABS.